\documentclass[aps,prd,nofootinbib,superscriptaddress,12pt]{revtex4}
\usepackage{amsmath}
\usepackage{graphicx}
\usepackage{dcolumn}
\usepackage{bm}
\usepackage{amssymb}
\usepackage{latexsym}

\newcommand{\be}{\begin{equation}}
\newcommand{\ee}{\end{equation}}
\newcommand{\bq}{\begin{eqnarray}}
\newcommand{\eq}{\end{eqnarray}}

\begin{document}

\title{Constraints on holographic dark energy from latest supernovae,
galaxy clustering, and cosmic microwave background anisotropy
observations}

\author{Xin Zhang}
\affiliation{National Astronomical Observatories, Chinese Academy of
Sciences, Beijing 100012, People's Republic of China} \affiliation{
Kavli Institute for Theoretical Physics China, Institute of
Theoretical Physics, Chinese Academy of Sciences (KITPC/ITP-CAS),
P.O.Box 2735, Beijing 100080, People's Republic of China}
\author{Feng-Quan Wu}
\affiliation{National Astronomical Observatories, Chinese Academy of
Sciences, Beijing 100012, People's Republic of China}

\begin{abstract}
The holographic dark energy model is proposed by Li as an attempt
for probing the nature of dark energy within the framework of
quantum gravity. The main characteristic of holographic dark energy
is governed by a numerical parameter $c$ in the model. The parameter
$c$ can only be determined by observations. Thus, in order to
characterize the evolving feature of dark energy and to predict the
fate of the universe, it is of extraordinary importance to constrain
the parameter $c$ by using the currently available observational
data. In this paper, we derive constraints on the holographic dark
energy model from the latest observational data including the gold
sample of 182 Type Ia supernovae (SNIa), the shift parameter of the
cosmic microwave background (CMB) given by the three-year {\it
Wilkinson Microwave Anisotropy Probe} ({\it WMAP}) observations, and
the baryon acoustic oscillation (BAO) measurement from the Sloan
Digital Sky Survey (SDSS). The joint analysis gives the fit results
in 1-$\sigma$: $c=0.91^{+0.26}_{-0.18}$ and $\Omega_{\rm m0}=0.29\pm
0.03$. That is to say, though the possibility of $c<1$ is more
favored, the possibility of $c>1$ can not be excluded in one-sigma
error range, which is somewhat different from the result derived
from previous investigations using earlier data. So, according to
the new data, the evidence for the quintom feature in the
holographic dark energy model is not as strong as before.

\end{abstract}

\pacs{98.80.-k, 95.36.+x}

\maketitle

\section{Introduction}\label{sec:intro}

Observations of Type Ia supernovae (SNIa) indicate that the universe
is experiencing an accelerating expansion at the present stage
\cite{Riess98,Perl99}. This cosmic acceleration has also been
confirmed by observations of large scale structure (LSS)
\cite{Tegmark04,Abazajian04} and measurements of the cosmic
microwave background (CMB) anisotropy \cite{Spergel03,Bennett03}.
The cause for this cosmic acceleration is usually referred to as
``dark energy'', a mysterious exotic matter with large enough
negative pressure, whose energy density has been a dominative power
of the universe (for reviews see e.g.
\cite{Weinberg89,SS00,Carroll01,Peebles03,Padmanabhan03,CST06}). The
astrophysical feature of dark energy is that it remains unclustered
at all scales where gravitational clustering of baryons and
nonbaryonic cold dark matter can be seen. Its gravity effect is
shown as a repulsive force so as to make the expansion of the
universe accelerate when its energy density becomes dominative power
of the universe. The combined analysis of cosmological observations
suggests that the universe is spatially flat, and consists of about
$70\%$ dark energy, $30\%$ dust matter (cold dark matter plus
baryons), and negligible radiation. Although we can affirm that the
ultimate fate of the universe is determined by the feature of dark
energy, the nature of dark energy as well as its cosmological origin
remain enigmatic at present. However, we still can propose some
candidates to interpret or describe the properties of dark energy.
The most obvious theoretical candidate of dark energy is the
cosmological constant $\lambda$ \cite{Einstein17} which always
suffers from the ``fine-tuning'' and ``cosmic coincidence'' puzzles.
Theorists have made lots of efforts to try to resolve the
cosmological constant problem, but all these efforts were turned out
to be unsuccessful. Numerous other candidates for dark energy have
also been proposed in the literature, such as an evolving canonical
scalar field
\cite{Peebles88,Wetterich88,Caldwell98,Zlatev99,Zhang05a,Zhang05b}
usually referred to as quintessence, the phantom energy
\cite{Caldwell02,Caldwell03,Kahya06} with an equation-of-state
smaller than $-1$ violating the weak energy condition, the quintom
energy \cite{Feng05,Guo05,Zhang05c,Zhangxf06,ZhangJ06,Cai07} with an
equation-of-state evolving across $-1$, the hessence model
\cite{Wei05a,Wei05b,Wei06}, the Chaplygin gas model
\cite{Kamenshchik01,Bento02,ZhangWZ06}, and so forth.

Actually, the dark energy problem may be in principle a problem
belongs to quantum gravity domain \cite{Witten00}. Another promising
model for dark energy, the holographic dark energy model, was
proposed by Li \cite{Li04} from some considerations of fundamental
principle in the quantum gravity. It is well known that the
holographic principle is an important result of the recent
researches for exploring the quantum gravity or string theory
\cite{'t Hooft93,Susskind94}. This principle is enlightened by
investigations of the quantum property of black holes. Roughly
speaking, in a quantum gravity system, the conventional local
quantum field theory will break down. The reason is rather simple:
For a quantum gravity system, the conventional local quantum field
theory contains too many degrees of freedom, and such many degrees
of freedom will lead to the formation of black hole so as to break
down the effectiveness of the quantum field theory.

For an effective field theory in a box of size $L$, with UV cut-off
$\Lambda$ the entropy $S$ scales extensively, $S\sim L^3\Lambda^3$.
However, the peculiar thermodynamics of black hole
\cite{Bekenstein73,Bekenstein74,Bekenstein81,Bekenstein94,Hawking75,Hawking76}
has led Bekenstein to postulate that the maximum entropy in a box of
volume $L^3$ behaves nonextensively, growing only as the area of the
box, i.e. there is a so-called Bekenstein entropy bound, $S\leq
S_{BH}\equiv\pi M_{\rm Pl}^2L^2$. This nonextensive scaling suggests
that quantum field theory breaks down in large volume. To reconcile
this breakdown with the success of local quantum field theory in
describing observed particle phenomenology, Cohen et al.
\cite{Cohen99} proposed a more restrictive bound --- the energy
bound. They pointed out that in quantum field theory a short
distance (UV) cut-off is related to a long distance (IR) cut-off due
to the limit set by forming a black hole. In other words, if the
quantum zero-point energy density $\rho_{\rm vac}$ is relevant to a
UV cut-off, the total energy of the whole system with size $L$
should not exceed the mass of a black hole of the same size, thus we
have $L^3\rho_{\rm vac}\leq LM_{\rm Pl}^2$. This means that the
maximum entropy is in order of $S_{BH}^{3/4}$. When we take the
whole universe into account, the vacuum energy related to this
holographic principle \cite{'t Hooft93,Susskind94} is viewed as dark
energy, usually dubbed holographic dark energy (its density is
denoted as $\rho_{\rm de}$ hereafter). The largest IR cut-off $L$ is
chosen by saturating the inequality so that we get the holographic
dark energy density
\begin{equation}
\rho_{\rm de}=3c^2M_{\rm Pl}^2L^{-2}~,\label{de}
\end{equation} where $c$ is a numerical constant, and $M_{\rm Pl}\equiv 1/\sqrt{8\pi
G}$ is the reduced Planck mass. If we take $L$ as the size of the
current universe, for instance the Hubble radius $H^{-1}$, then the
dark energy density will be close to the observational result.
However, Hsu \cite{Hsu04} pointed out that this yields a wrong
equation of state for dark energy. Li \cite{Li04} subsequently
proposed that the IR cut-off $L$ should be taken as the size of the
future event horizon
\begin{equation}
R_{\rm eh}(a)=a\int_t^\infty{dt'\over a(t')}=a\int_a^\infty{da'\over
Ha'^2}~.\label{eh}
\end{equation} Then the problem can be solved nicely and the
holographic dark energy model can thus be constructed successfully.
The holographic dark energy scenario may provide simultaneously
natural solutions to both dark energy problems as demonstrated in
\cite{Li04}. For extensive studies on the holographic dark energy
model see e.g.
\cite{HuangLi04,Enqvist04,Ke05,HuangLi05,Zhang05d,Pavon05,Wangb05,
Kim05,Nojiri06,Hu06,Zhang06a, Zhang06b,Chen06}.

The holographic dark energy model has been tested and constrained by
various astronomical observations, such as SNIa \cite{HuangGong04},
CMB \cite{Enqvist05,Shen05,Kao05}, combination of SNIa, CMB and LSS
\cite{ZhangWu05}, the X-ray gas mass fraction of galaxy clusters
\cite{ChangWZ06}, and the differential ages of passively evolving
galaxies \cite{Yi06}. Recently, the three-year data of {\it
Wilkinson Microwave Anisotropy Probe} ({\it WMAP}) observations
\cite{Spergel06} were announced. Moreover, Riess et al.
\cite{Riess06} lately released the up-to-date 182 ``gold'' data of
SNIa from various sources analyzed in a consistent and robust mannor
with reduced calibration errors arising from systematics. This paper
aims at placing new observational constraints on the holographic
dark energy model by using the gold sample of 182 SNIa compiled by
Riess et al. \cite{Riess06}, the CMB shift parameter derived from
three-year {\it WMAP} observations \cite{Wang06}, and the baryon
acoustic oscillations detected in the large-scale correlation
function of Sloan Digital Sky Survey (SDSS) luminous red galaxies
\cite{Eisenstein05}.

This paper is organized as follows: In section \ref{sec:holode} we
discuss the basic characteristics of the holographic dark energy
model. In section \ref{sec:constrain}, we perform constraints on the
holographic dark energy model by using the up-to-date observational
datasets. Finally, we give the concluding remarks in section
\ref{sec:concl}.

\section{The model of holographic dark energy}\label{sec:holode}

In this section, we shall review the holographic dark energy model
briefly and discuss the basic characteristics of this model. Now let
us consider a spatially flat Friedmann-Robertson-Walker (FRW)
universe with matter component $\rho_{\rm m}$ (including both baryon
matter and cold dark matter) and holographic dark energy component
$\rho_{\rm de}$, the Friedmann equation reads
\begin{equation}
3M_{\rm Pl}^2H^2=\rho_{\rm m}+\rho_{\rm de}~,
\end{equation} or equivalently,
\begin{equation}
E(z)\equiv {H(z)\over H_0}=\left(\Omega_{\rm m0}(1+z)^3\over
1-\Omega_{\rm de}\right)^{1/2},\label{Ez}
\end{equation} where $z=(1/a)-1$ is the redshift of the universe.
Note that we always assume spatial flatness throughout this paper as
motivated by inflation. Combining the definition of the holographic
dark energy (\ref{de}) and the definition of the future event
horizon (\ref{eh}), we derive
\begin{equation}
\int_a^\infty{d\ln a'\over Ha'}={c\over Ha\sqrt{\Omega_{\rm
de}}}~.\label{rh}
\end{equation} We notice that the Friedmann
equation (\ref{Ez}) implies
\begin{equation}
{1\over Ha}=\sqrt{a(1-\Omega_{\rm de})}{1\over H_0\sqrt{\Omega_{\rm
m0}}}~.\label{fri}
\end{equation} Substituting (\ref{fri}) into (\ref{rh}), one
obtains the following equation
\begin{equation}
\int_x^\infty e^{x'/2}\sqrt{1-\Omega_{\rm de}}dx'=c
e^{x/2}\sqrt{{1\over\Omega_{\rm de}}-1}~,
\end{equation} where $x=\ln a$. Then taking derivative with respect to $x$ in both
sides of the above relation, we get easily the dynamics satisfied by
the dark energy, i.e. the differential equation about the fractional
density of dark energy,
\begin{equation}
\Omega'_{\rm de}=-(1+z)^{-1}\Omega_{\rm de}(1-\Omega_{\rm
de})\left(1+{2\over c}\sqrt{\Omega_{\rm de}}\right),\label{deq}
\end{equation}
where the prime denotes the derivative with respect to the redshift
$z$. This equation describes behavior of the holographic dark energy
completely, and it can be solved exactly \cite{Li04}. From the
energy conservation equation of the dark energy, the equation of
state of the dark energy can be given \cite{Li04}
\begin{equation}
w=-1-{1\over 3}{d\ln\rho_{\rm de}\over d\ln a}=-{1\over 3}(1+{2\over
c}\sqrt{\Omega_{\rm de}})~.\label{w}
\end{equation} Note that the formula
$\rho_{\rm de}={\Omega_{\rm de}\over 1-\Omega_{\rm de}}\rho_{\rm
m0}a^{-3}$ and the differential equation of $\Omega_{\rm de}$
(\ref{deq}) are used in the second equal sign. It can be seen
clearly that the equation of state of the holographic dark energy
evolves dynamically and satisfies $-(1+2/c)/3\leq w\leq -1/3$ due to
$0\leq\Omega_{\rm de}\leq 1$. Hence, we see clearly that when taking
the holographic principle into account the vacuum energy becomes
dynamically evolving dark energy.

The parameter $c$ plays a significant role in this model. If one
takes $c=1$, the behavior of the holographic dark energy will be
more and more like a cosmological constant with the expansion of the
universe, such that ultimately the universe will enter the de Sitter
phase in the far future. As is shown in \cite{Li04}, if one puts the
parameter $\Omega_{\rm de0}=0.73$ into (\ref{w}), then a definite
prediction of this model, $w_0=-0.903$, will be given. On the other
hand, if $c<1$, the holographic dark energy will exhibit appealing
behavior that the equation of state crosses the
``cosmological-constant boundary'' (or ``phantom divide'') $w=-1$
during the evolution. This kind of dark energy is referred to as
``quintom'' \cite{Feng05} which is slightly favored by current
observations, see e.g.
\cite{Alam04,Huterer05,Zhao05,Xia06,Zhao06a,Li06,Gong06,Zhao06b}. If
$c>1$, the equation of state of dark energy will be always larger
than $-1$ such that the universe avoids entering the de Sitter phase
and the Big Rip phase. Hence, we see explicitly, the value of $c$ is
very important for the holographic dark energy model, which
determines the feature of the holographic dark energy as well as the
ultimate fate of the universe. For an illustrative example, see
Figure 1 in \cite{ZhangWu05}, in which the selected evolutions in
different $c$ for the equation of state of holographic dark energy
are plotted. It is clear to see that the cases in $c\geq 1$ always
evolve in the region of $w\geq -1$, whereas the case of $c<1$
behaves as a quintom whose equation of state $w$ crosses the
cosmological constant boundary $w=-1$ during the evolution. It has
been shown in previous analyses of observational data
\cite{ZhangWu05,ChangWZ06,Yi06} that the holographic dark energy
exhibits quintom-like behavior basically within statistical error
one sigma.

\begin{figure}[htbp]
\begin{center}
\includegraphics[scale=1.2]{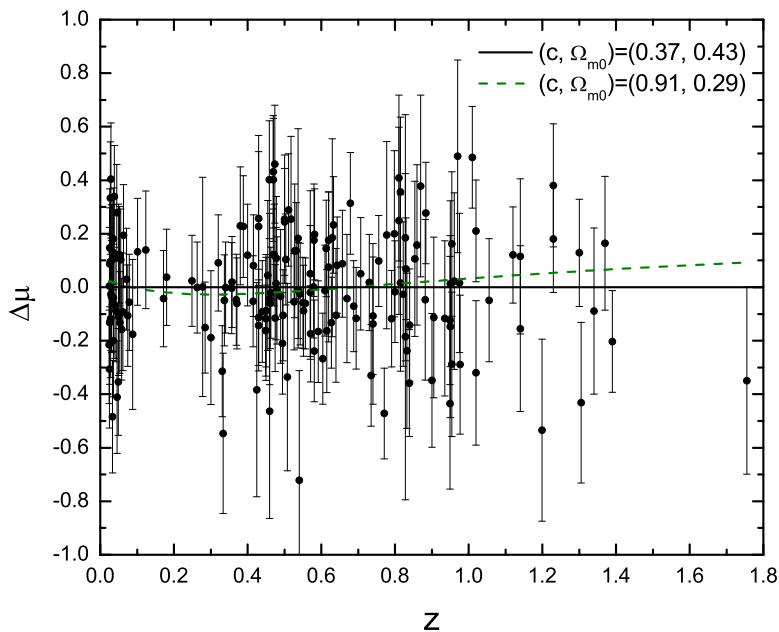}
\caption{Gold sample of 182 SNIa residual Hubble diagram comparing
holographic dark energy model with best-fit values for parameters.
The black solid line represents the best-fit for SNIa alone analysis
with $(c,~\Omega_{\rm m0})=(0.37,~0.43)$; The green dashed line
represents the best-fit for SNIa+CMB+LSS joint analysis with
$(c,~\Omega_{\rm m0})=(0.91,~0.29)$. Data and model are shown
relative to the case of $(c,~\Omega_{\rm
m0})=(0.37,~0.43)$.}\label{fig:hubble}
\end{center}
\end{figure}

\begin{figure}[htbp]
\begin{center}
\includegraphics[scale=1.2]{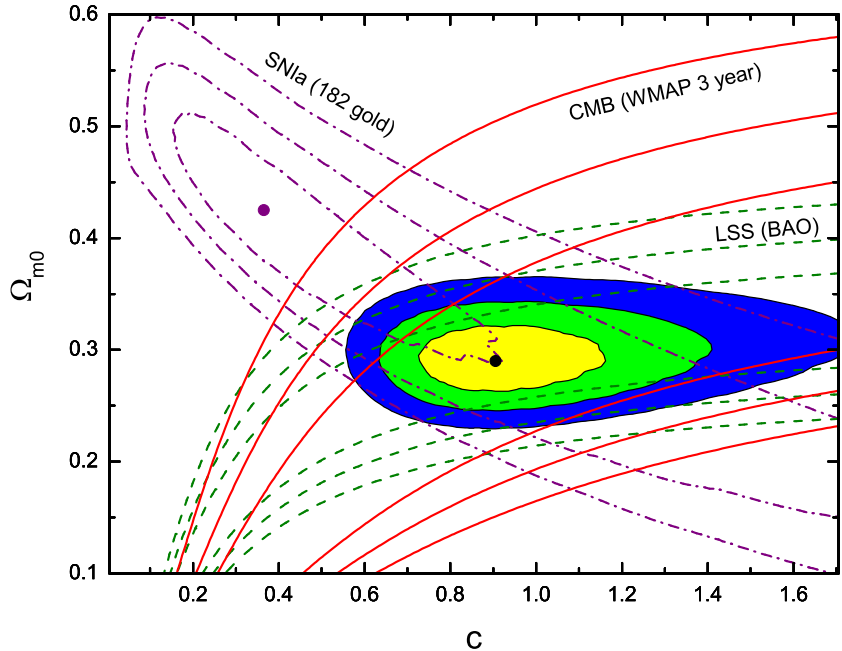}
\caption{Probability contours at $68.3\%$, $95.4\%$, and $99.7\%$
confidence levels in $(c,~\Omega_{\rm m0})$-plane, for the
holographic dark energy model, from the gold sample of SNIa data
(purple dot-dashed contours), from the shift parameter $R$ in CMB
(red solid contours), from the parameter $A$ in BAO found in the
SDSS (green dashed contours), and from the combination of the three
databases (shaded contours). The points show the best-fit cases for
SNIa alone analysis and for SNIa+CMB+LSS joint analysis,
respectively.}\label{fig:contour}
\end{center}
\end{figure}

\begin{figure}[htbp]
\begin{center}
\includegraphics[scale=1.2]{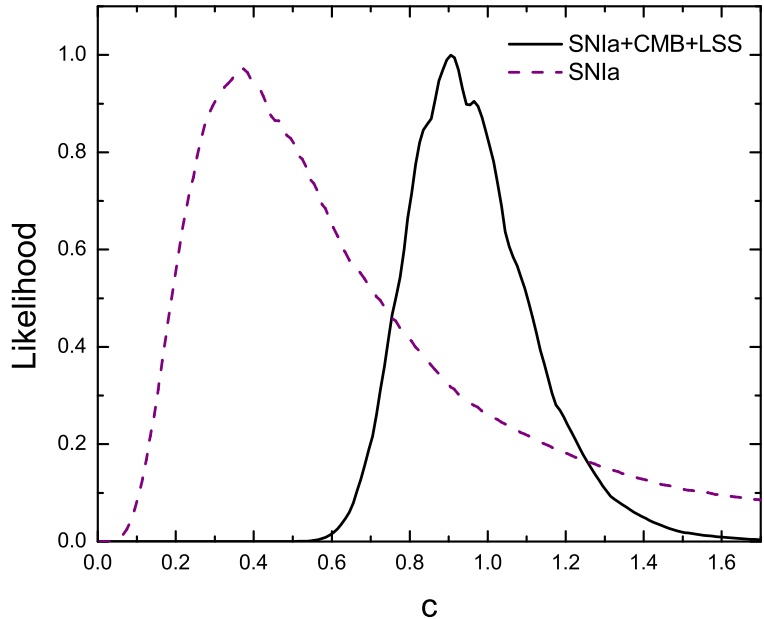}
\caption{Probability distribution of parameter $c$ for the fits of
SNIa alone and SNIa+CMB+LSS.}\label{fig:likelihood}
\end{center}
\end{figure}

\section{Constraints from latest SNIa,
LSS, and CMB observations}\label{sec:constrain}

In this section, we constrain the parameters in the holographic dark
energy model and analyze the evolutionary behavior of holographic
dark energy by using the latest observational data of SNIa combined
with the information from CMB and LSS observations.

Recently, the up-to-date gold sample of SNIa consists of 182 data
was compiled by Riess et al. \cite{Riess06}. It contains 119 points
from the previous sample compiled in \cite{Riess04} and 16 points
with $0.46<z<1.39$ discovered recently \cite{Riess06} by the {\it
Hubble Space Telescope} ({\it HST}). It also incorporates 47 points
($0.25<z<0.96$) from the first year release of the Supernova Legacy
Survey (SNLS) sample \cite{Astier06} out of a total of 73 distant
SNIa. Some previous gold data were excluded in \cite{Riess06} due to
highly uncertain color measurements, high extinction $A_V>0.5$ and a
redshift cut $z<0.0233$, to avoid the influence of a possible local
``Hubble Bubble'', so as to define a high-confidence sample. The
total gold sample spans a wide range of redshift $0.024<z<1.76$. For
recent usages of the new sample, see e.g.
\cite{Barger06,Alam06,Li06,Gong06,ZhangJ06,Nesseris06,Zhao06b,Wei06,Dutta06}.
We shall analyze the holographic dark energy model in the light of
the new gold sample of SNIa.

The SNIa observations directly measure the apparent magnitude $m$ of
a supernova and its redshift $z$. The apparent magnitude $m$ is
related to the luminosity distance $d_L$ of the supernova through
\begin{equation}
m(z)=M+5\log_{10}(d_L(z)/{\rm Mpc})+25~,
\end{equation} where $M$ is the absolute magnitude which is
believed to be constant for all Type Ia supernovae, and the
luminosity distance-redshift relation is
\begin{equation}
d_L(z)=\left({{\cal L}\over 4\pi{\cal
F}}\right)^{1/2}=H_0^{-1}(1+z)\int_0^z{dz'\over E(z')}~,
\end{equation} where ${\cal L}$ is the absolute luminosity which
is a known value for the standard candle SNIa, $\cal F$ is the
measured flux, $H_0^{-1}$ (here we use the natural unit, namely the
speed of light is defined to be 1) represents the Hubble distance
with value $H_0^{-1}=2997.9h^{-1}$ Mpc, and $E(z)=H(z)/H_0$ is
expressed in equation (\ref{Ez}). Note that the dynamical behavior
of $\Omega_{\rm de}$ is governed by differential equation
(\ref{deq}). In order to place constraints on the holographic dark
energy model, we perform $\chi^2$ statistics for the model
parameters $(c, \Omega_{\rm m0})$ and the present Hubble parameter
$H_0$. For the SNIa analysis, we have
\begin{equation}
\chi^2_{\rm SN}=\sum\limits_{i=1}^{182}{[\mu_{\rm obs}(z_i)-\mu_{\rm
th}(z_i)]^2\over \sigma_i^2}~,\label{chisn}
\end{equation} where the extinction-corrected distance moduli
$\mu(z)$ is defined as $\mu(z)=m(z)-M$, and $\sigma_i$ is the total
uncertainty in the observation. The likelihood ${\cal L}\propto
e^{-\chi^2/2}$ if the measurement errors are Gaussian. The best-fit
for the analysis of gold sample of 182 SNIa happens at $c=0.37$,
$\Omega_{\rm m0}=0.43$, and $h=0.64$, with $\chi^2_{\rm
min}=156.60$. The gold sample is illustrated on a residual Hubble
diagram with respect to our best-fit universe in Figure
\ref{fig:hubble}. Since we concentrate on the model parameters $(c,
\Omega_{\rm m0})$, we need to marginalize over the Hubble parameter
$H_0$. Note that marginalizing over $H_0$ is equivalent to
evaluating $\chi^2$ at its minimum with respect to $H_0$
\cite{Barris04}. We marginalize over the nuisance parameter $h$ and
show the probability contours at $68.3\%$, $95.4\%$, and $99.7\%$
confidence levels for $c$ vs. $\Omega_{\rm m0}$ in Figure
\ref{fig:contour}, from the constraints of gold sample of SNIa, as
purple dot-dashed contours. The 1 $\sigma$ fit values for the model
parameters are: $c=0.37^{+0.56}_{-0.21}$ and $\Omega_{\rm
m0}=0.43^{+0.08}_{-0.14}$. We see that the parameter $c$ in 1
$\sigma$ range, $0.16<c<0.93$, is smaller than 1, making the
holographic dark energy behave as quintom with equation-of-state
evolving across $w=-1$, according to this analysis.

On the other hand, the above analysis shows that the SNIa data alone
seem not sufficient to constrain the holographic dark energy model
strictly. The confidence region of $c-\Omega_{\rm m0}$ plane is
rather large, especially for the parameter $c$. Moreover, it is
remarkable that the best fit value of $\Omega_{\rm m0}$ of this
model is evidently larger than that of the $\Lambda$CDM model. For
comparison, we refer to the {\it WMAP} result for $\Omega_{\rm m0}$
in $\Lambda$CDM model: $\Omega_{\rm m0}=0.24^{+0.03}_{-0.04}$
\cite{Spergel06}. As has been elucidated in \cite{ZhangWu05} that
for the holographic dark energy the fit of SNIa data is very
sensitive to the Hubble parameter $H_0$, so it is very important to
find other observational quantities irrelevant to $H_0$ as a
complement to SNIa data. Fortunately, such suitable data can be
found in the probes of CMB and LSS.

\begin{table*}
\caption{\label{tab:para} Constraints from observational data. The
fit values of $c$ and $\Omega_{\rm m0}$ are given in 1-$\sigma$
errors; the fit value of $h$ is given at best-fit case; the value of
$\chi^2_{\rm min}$ is also for best-fit.} \footnotesize
\begin{center}
\begin{tabular}{ccc}
\hline\hline \\
Parameter/Quantity ~~& ~~SNIa gold sample alone ~~&~~ SNIa+CMB+LSS\\
 \hline \\
$c\dotfill$&$0.37^{+0.56}_{-0.21}$   &$0.91^{+0.26}_{-0.18}$   \\
$\Omega_{\rm m0}\dotfill$&$0.43^{+0.08}_{-0.14}$  &$0.29^{+0.03}_{-0.03}$  \\
$h\dotfill$&$0.64$  &$0.63$  \\
$\chi^2_{\rm min}\dotfill$&$156.60$  &$158.97$  \\
\hline
\end{tabular}
\end{center}
\end{table*}

For the CMB data, we use the CMB shift parameter. The CMB shift
parameter $R$ is perhaps the least model-independent parameter that
can be extracted from CMB data. The shift parameter $R$ is given by
\cite{Bond97}
\begin{equation}
R\equiv \Omega_{\rm m0}^{1/2}\int_0^{z_{\rm CMB}}{dz'\over E(z')},
\end{equation} where $z_{\rm CMB}=1089$ is the redshift of recombination.
The value of the shift parameter $R$ can be determined by three-year
integrated {\it WMAP} analysis \cite{Spergel06}, and has been
updated by \cite{Wang06} to be $1.70\pm 0.03$ independent of the
dark energy model. For the LSS data, we use the measurement of the
BAO peak in the distribution of SDSS luminous red galaxies (LRGs).
The SDSS BAO measurement \cite{Eisenstein05} gives
$A=0.469(n_s/0.98)^{-0.35}\pm 0.017$ (independent of a dark energy
model) at $z_{\rm BAO}=0.35$, where $A$ is defined as
\begin{equation}
A\equiv \Omega_{\rm m0}^{1/2} E(z_{\rm BAO})^{-1/3}\left[{1\over
z_{\rm BAO}}\int_0^{z_{\rm BAO}}{dz'\over E(z')}\right]^{2/3}.
\end{equation}
Here the scalar spectral index is taken to be $n_s=0.95$ as measured
by the three-year {\it WMAP} data \cite{Spergel06}. We notice that
both $R$ and $A$ are independent of $H_0$; thus these quantities can
provide robust constraint as complement to SNIa data on the
holographic dark energy model.

\begin{figure}[htbp]
\begin{center}
\includegraphics[scale=1.2]{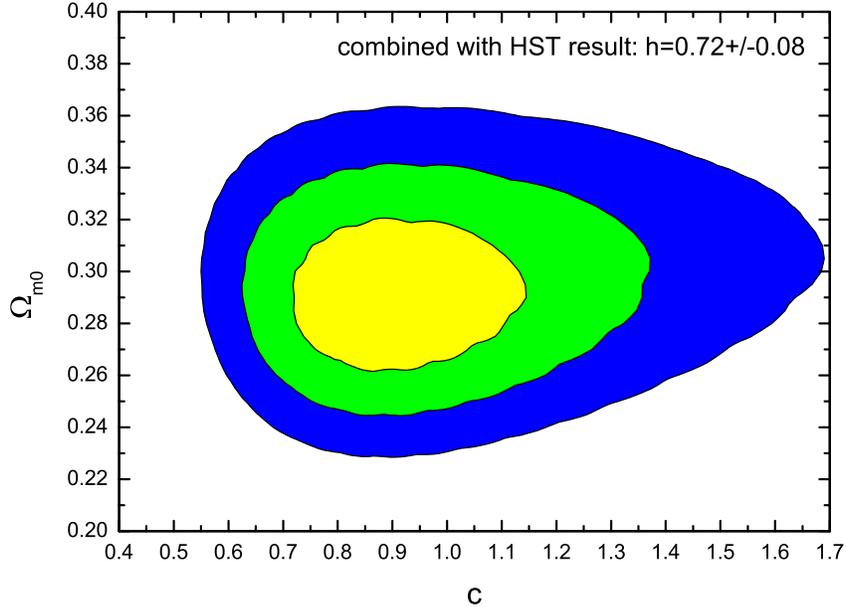}
\caption{Probability contours at $68.3\%$, $95.4\%$, and $99.7\%$
confidence levels in $(c,~\Omega_{\rm m0})$-plane, for the
holographic dark energy model, from the constraints of the
combination of SNIa, CMB and LSS. In this analysis, the {\it HST}
prior $h=0.72\pm 0.08$ is also considered by adding a term
$\chi^2_{HST}=[(h-0.72)/0.08]^2$ to the total $\chi^2$. The fit
values for model parameters with one-sigma errors are
$c=0.91^{+0.23}_{-0.19}$ and $\Omega_{\rm m0}=0.29\pm 0.03$, which
are almost the same as the results without prior, see Figure 2.
}\label{fig:hstprior}
\end{center}
\end{figure}

We now perform a combined analysis of SNIa, CMB, and LSS on the
constraints of the holographic dark energy model. We use the
$\chi^2$ statistics
\begin{equation}
\chi^2=\chi_{\rm SN}^2+\chi_{\rm CMB}^2+\chi_{\rm LSS}^2~,
\end{equation} where $\chi_{\rm SN}^2$ is given by equation
(\ref{chisn}) for SNIa statistics, $\chi_{\rm CMB}^2=[({R}-{R}_{\rm
obs})/\sigma_{R}]^2$ and $\chi_{\rm LSS}^2=[(A-A_{\rm
obs})/\sigma_A]^2$ are contributions from CMB and LSS data,
respectively. The main results are shown in Figure
\ref{fig:contour}. In this figure, we show the contours of $68.3\%$,
$95.4\%$, and $99.7\%$ confidence levels in the $c-\Omega_{\rm m0}$
plane. The constraints from the shift parameter $R$ in CMB are
illustrated by the red solid contours; the constraints from the
parameter $A$ in BAO are illustrated by green dashed contours; the
joint constraints from SNIa+CMB+LSS are shown as shaded contours. It
is clear to see that the combined analysis of SNIa, CMB and LSS data
provides a fairly tight constraint on the holographic dark energy
model, comparing to the constraint from the SNIa gold sample alone.
The fit values for the model parameters with 1-$\sigma$ errors are
$c=0.91^{+0.26}_{-0.18}$ and $\Omega_{\rm m0}=0.29\pm 0.03$ with
$\chi^2_{\rm min}=158.97$. For comparison, the fit results from the
SNIa gold sample alone and from the combination of SNIa, CMB and LSS
are shown in Table \ref{tab:para}. We also show the best-fit case of
SNIa+CMB+LSS analysis on the residual Hubble diagram with respect to
the best-fit case of SNIa alone analysis in Figure \ref{fig:hubble}.
We see clearly that in the joint analysis the derived value for
matter density $\Omega_{\rm m0}$ is very reasonable. In addition, it
should be emphasized that what is of importance for this model is
the determination of the value of $c$. In Figure
\ref{fig:likelihood} we plot the 1-dimensional likelihood function
for $c$, marginalizing over the other parameters. We notice that the
best-fit value of $c$ in this analysis is enhanced to around 0.91.
Intriguingly, the range of $c$ in 1-$\sigma$ error, $0.73<c<1.17$,
is not capable of ruling out the probability of $c>1$; this
conclusion is somewhat different from those derived from previous
investigations using earlier data. In previous work, for instance,
\cite{ZhangWu05} and \cite{ChangWZ06}, the 1-$\sigma$ range of $c$
obtained can basically exclude the probability of $c>1$ giving rise
to the quintessence-like behavior, supporting the quintom-like
behavior evidently. Though the present result (in 1-$\sigma$ error
range) from the analysis of the up-to-date observational data does
not support the quintom-like feature as strongly as before, the
best-fit value ($c=0.91$) still exhibits the holographic quintom
characteristic.

\begin{figure}[htbp]
\begin{center}
\includegraphics[scale=1.2]{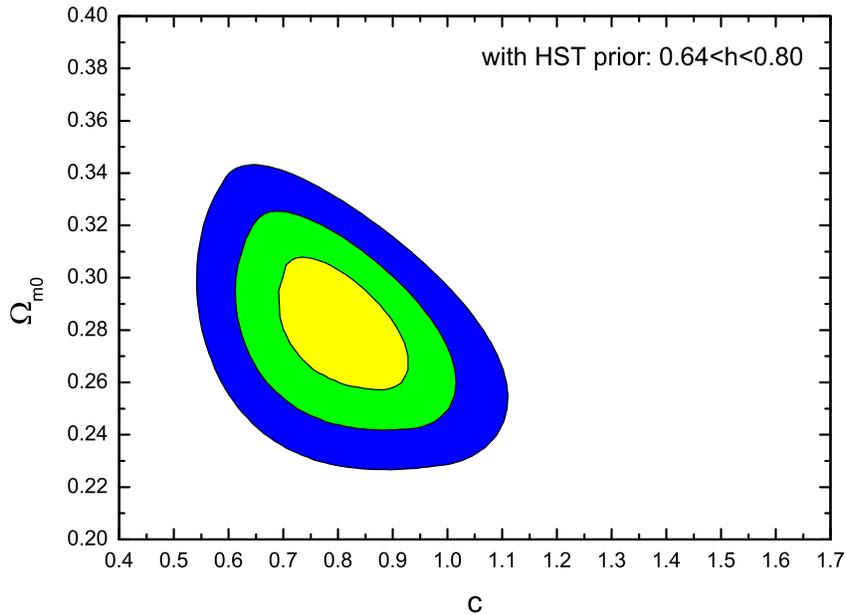}
\caption{Probability contours at $68.3\%$, $95.4\%$, and $99.7\%$
confidence levels in $(c,~\Omega_{\rm m0})$-plane, for the
holographic dark energy model, from the constraints of the
combination of SNIa, CMB and LSS, also with {\it HST} prior
$0.64<h<0.80$. The fit values for model parameters with one-sigma
errors are $c=0.82^{+0.11}_{-0.13}$ and $\Omega_{\rm
m0}=0.28^{+0.03}_{-0.02}$.}\label{fig:modestprior}
\end{center}
\end{figure}

\begin{figure}[htbp]
\begin{center}
\includegraphics[scale=1.2]{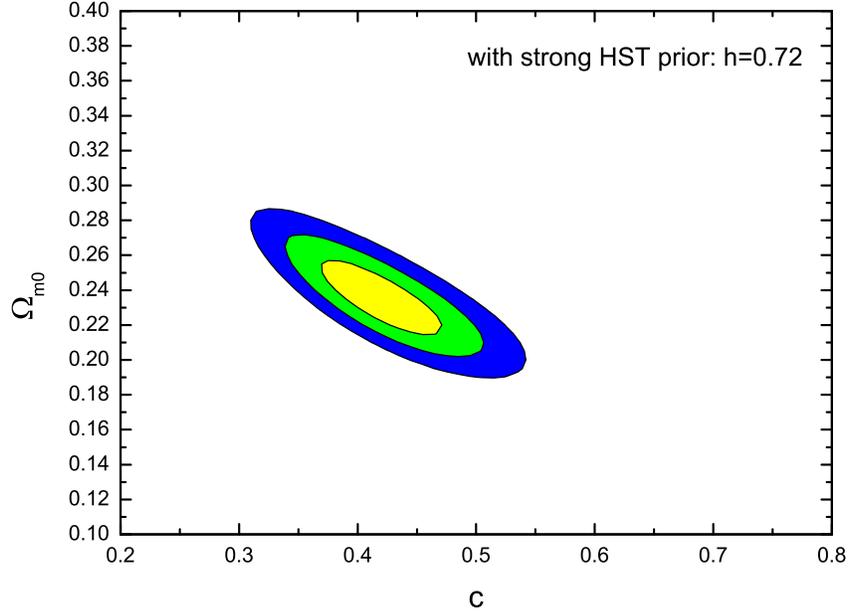}
\caption{Same as Figure 5,
but with strong-{\it HST} prior $h=0.72$. The fit values for model
parameters with one-sigma errors are $c=0.42\pm 0.05$ and
$\Omega_{\rm m0}=0.24^{+0.02}_{-0.03}$.}\label{fig:strongprior}
\end{center}
\end{figure}

Another problem of concern is that both the SNIa alone analysis and
the SNIa + CMB + LSS joint analysis predict a low value of
dimensionless Hubble constant $h$. For the Hubble constant, one of
the most reliable results comes from the {\it Hubble Space
Telescope} Key Project \cite{Freeman01}. This group has used the
empirical period-luminosity relations for Cepheid variable stars to
obtain distances to 31 galaxies, and calibrated a number of
secondary distance indicators measured over distances of 400 to 600
Mpc. The result they obtained is $h=0.72\pm 0.08$. It is remarkable
that, intriguingly, this result is in such good agreement with the
result derived from the {\it WMAP} CMB measurements,
$h=0.73^{+0.03}_{-0.04}$ (it should be pointed out that this result
is derived from a flat $\Lambda$CDM model assumption)
\cite{Spergel06}.

\begin{figure}[htbp]
\begin{center}
\includegraphics[scale=1.2]{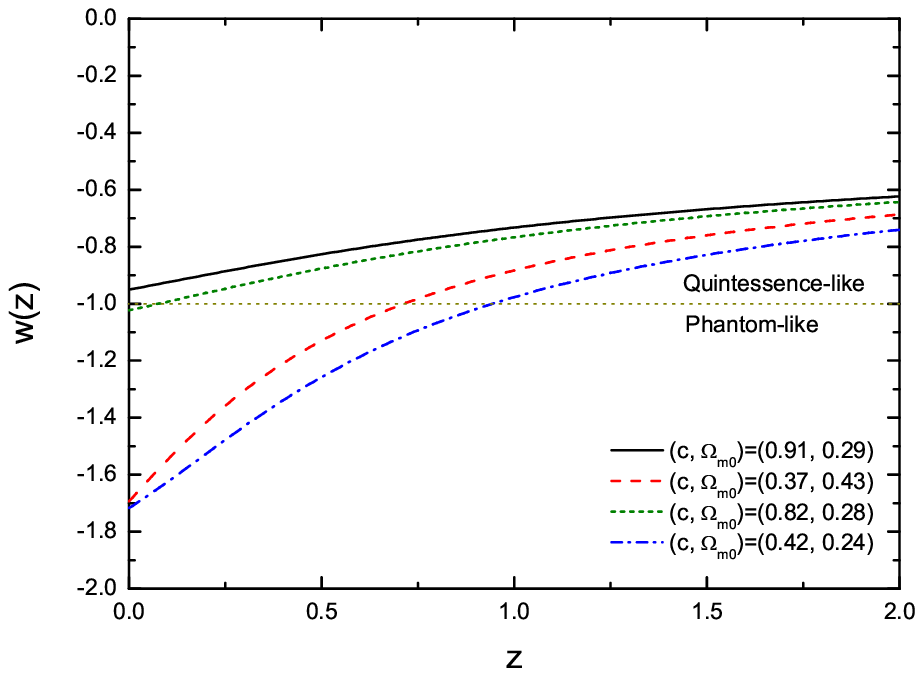}
\caption{The equation-of-state parameter of dark energy $w$ versus
redshift $z$, from some best-fit values of the holographic dark
energy model. }\label{fig:wz}
\end{center}
\end{figure}

\begin{figure}[htbp]
\begin{center}
\includegraphics[scale=1.2]{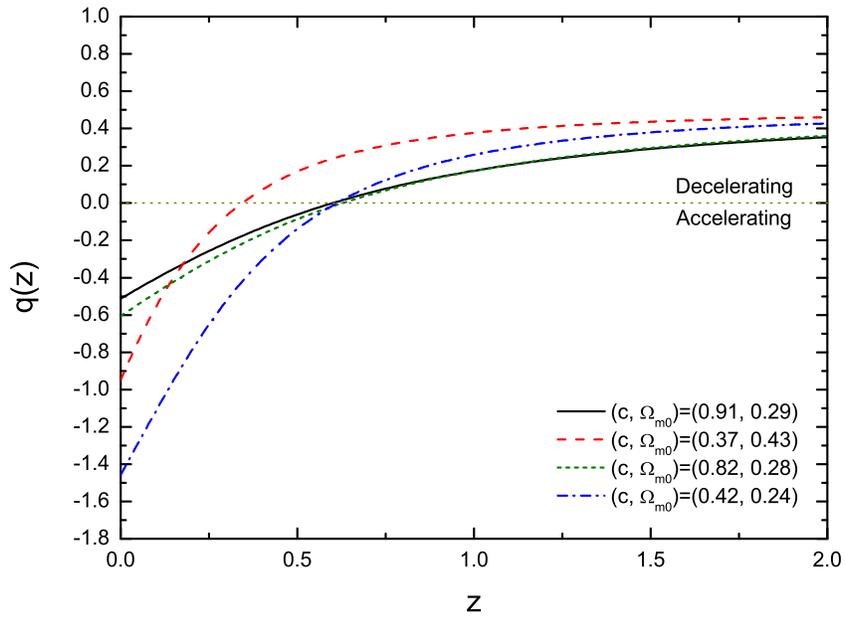}
\caption{Deceleration parameter $q$ versus redshift $z$, from some
best-fit values of the holographic dark energy model.}\label{fig:qz}
\end{center}
\end{figure}

In the follows we shall incorporate the {\it HST} Hubble constant
result into the SNIa + CMB + LSS data fitting. First, we add the
{\it HST} term, $\chi^2_{HST}=[(h-0.72)/0.08]^2$, to the total
$\chi^2$. The results we derived are shown in Firgure
\ref{fig:hstprior}, the 1, 2 and 3 $\sigma$ contours of confidence
levels in $c-\Omega_{\rm m0}$ plane. The fit values for model
parameters with one-sigma errors are $c=0.91^{+0.23}_{-0.19}$ and
$\Omega_{\rm m0}=0.29\pm 0.03$, which are almost the same as the
results from without {\it HST} data. So, we next take another way of
incorporating the {\it HST} prior, $0.64<h<0.80$, into account, in
the data analysis. When considering this prior, the confidence level
contours get shrinkage and left-shift in the $c-\Omega_{\rm m0}$
parameter-plane, as shown in Figure \ref{fig:modestprior}. In this
case the fit values for model parameters with one-sigma errors are
$c=0.82^{+0.11}_{-0.13}$ and $\Omega_{\rm m0}=0.28^{+0.03}_{-0.02}$.
We see that the holographic dark energy features quintom dark energy
within one-sigma range in this case. Furthermore, we also consider a
strong {\it HST} prior, fixing $h=0.72$, in order to see how
strongly biased constraints can be derived from a factitious prior
on $h$. We plot the results of this case in Figure
\ref{fig:strongprior}. The fit values for model parameters with
one-sigma errors are $c=0.42\pm 0.05$ and $\Omega_{\rm
m0}=0.24^{+0.02}_{-0.03}$. We find that the shrinkage and the
left-shift for the confidence level contours become more evident.
For illustrating the cosmological consequences led by the
observational constraints, we show the evolution cases of the
equation-of-state parameter $w(z)$ and the deceleration parameter
$q(z)$ according to some best-fit values of parameters of the
holographic dark energy model in Figure \ref{fig:wz} and
\ref{fig:qz}. The quintom feature with $w=-1$ crossing
characteristic for the holographic dark energy model can be easily
seen.

\section{Concluding remarks}\label{sec:concl}

The cosmic acceleration observed by distance-redshift relation
measurement of SNIa strongly supports the existence of dark energy.
The fantastic physical property of dark energy not only drives the
current cosmic acceleration, but also determines the ultimate fate
of the universe. However, hitherto, the nature of dark energy as
well as its cosmological origin still remain enigmatic for us.
Though the underlying theory of dark energy is still far beyond our
knowledge, it is guessed that the quantum gravity theory shall play
a significant role in resolving the dark energy enigma. The
holographic dark energy model is proposed as an attempt for probing
the nature of dark energy within the framework of quantum gravity,
i.e. it is based upon an important fundamental principle of quantum
gravity
--- holographic principle, so it possesses some significant features
of an underlying theory of dark energy. The main characteristic of
holographic dark energy is governed by a numerical parameter $c$ in
the model. This parameter, $c$, can only be determined by
observations. Hence, in order to characterize the evolving feature
of dark energy and to predict the fate of the universe, it is of
extraordinary importance to constrain the parameter $c$ by using the
currently available observational data.

In this paper, we have analyzed the holographic dark energy model by
using the up-to-date gold SNIa sample, combined with the CMB and LSS
data. Since the SNIa data are sensitive to the Hubble constant
$H_0$, while the shift parameter in CMB and the parameter in the BAO
are irrelevant to the Hubble parameter, the combination of these
datasets leads to strong constraints on the model parameters, as
shown in Figure \ref{fig:contour}. The joint analysis indicates
that, though the possibility of $c<1$ is more favored, the
possibility of $c>1$ can not be excluded in one-sigma error range,
which is somewhat different from the result derived from previous
investigations using earlier data (such as \cite{ZhangWu05}, in
which the result of $c<1$ is basically favored in 1-$\sigma$ range).
That is to say, according to the new data, the evidence for the
quintom feature in the holographic dark energy model is not as
strong as before. However, when considering the {\it HST} prior,
$0.64<h<0.80$, the quintom-like behavior can be supported in
one-sigma error range, as shown in Figure \ref{fig:modestprior}. On
the whole, the current observational data have no ability to
constrain the parameters in the holographic dark energy model on a
high precision level. We expect that the future high-precision
observations such as the SuperNova/Acceleration Project (SNAP) will
be capable of determining the value of $c$ exactly and thus
revealing the property of the holographic dark energy.

\section*{Acknowledgements}
We would like to thank Qing-Guo Huang, Miao Li, and Hao Wei for
helpful discussions. This work was supported by the National Natural
Science Foundation of China, the K. C. Wong Education Foundation
(Hong Kong), and the Postdoctoral Science Foundation of China.


\end{document}